\documentclass{article}

\usepackage{amstext,amsmath,amssymb}
\usepackage[dvips]{color}
\usepackage{graphicx}
\usepackage{epsfig}
\usepackage{cite}
\usepackage[left=4cm,top=3cm,bottom=3cm,right=4cm]{geometry}

\begin{document}

\numberwithin{equation}{section}

\setcounter{secnumdepth}{2}

\title{Explicit factorization of external coordinates in constrained Statistical Mechanics models}

\author{Pablo Echenique$^{1,2}$\footnote{Corresponding author. E-mail
address: {\tt pnique@unizar.es}}\ \ and Iv\'{a}n Calvo$^{1,2}$
\vspace{0.4cm}\\ $^{1}$ {\small Departamento de F\'{\i}sica
Te\'orica, Facultad de Ciencias, Universidad de Zaragoza,}\\
{\small Pedro Cerbuna 12, 50009, Zaragoza, Spain.}\\
$^{2}$ {\small Instituto de Biocomputaci\'on y
F\'{\i}sica de los Sistemas Complejos (BIFI),}\\ {\small Edificio
Cervantes, Corona de Arag\'on 42, 50009, Zaragoza, Spain.}}

\date{\today}

\maketitle

\begin{abstract}

If a macromolecule is described by curvilinear coordinates or rigid
constraints are imposed, the equilibrium probability density that must
be sampled in Monte Carlo simulations includes the determinants of
different mass-metric tensors. In this work, we explicitly write the
determinant of the mass-metric tensor $G$ and of the reduced
mass-metric tensor $g$, \emph{for any molecule, general internal
coordinates and arbitrary constraints}, as a product of two functions;
one depending only on the external coordinates that describe the
overall translation and rotation of the system, and the other only on
the internal coordinates. This work extends previous results in the
literature, proving with full generality that one may integrate out
the external coordinates and perform Monte Carlo simulations in the
internal conformational space of macromolecules. In addition, we give
a general mathematical argument showing that the factorization is a
consequence of the symmetries of the metric tensors involved. Finally,
the determinant of the mass-metric tensor $G$ is computed explicitly
in a set of curvilinear coordinates specially well-suited for general
branched molecules.
\vspace{0.2cm}\\ {\bf PACS:} 05.20.-y, 36.10.-k, 87.14.-g, 87.15.-v, 87.15.Aa, 89.75.-k
\vspace{0.2cm}\\

\end{abstract}

\section{Introduction}
\label{sec:introduction}

Monte Carlo simulations are among the most useful tools for studying
the behavior of macromolecules in thermal equilibrium
\cite{PE:Che2005JCP,PE:She2005BPJ,PE:Kli2004JCP,PE:Niv2004JMB,PE:Shi2002PNAS,PE:Han1999COSB,PE:Sen1998JCC,PE:Alm1990MP}. Typically,
the simulations are carried out in the \emph{coordinate space}, i.e.,
the momenta are averaged out and Monte Carlo movements that only
change the coordinates of the system are designed.

Also, the most interesting properties of macromolecules depend only on
conformational transitions in the internal subspace of the whole
coordinate space. The protein folding problem
\cite{PE:Alo2004BOOK,PE:Dob2003NAT,PE:Plo2000JCP,PE:Dil1999PSC}, the
docking of ligands to proteins \cite{PE:Tay2002JCAMD}, or proteins to
proteins \cite{PE:Smi2002COSB}, the prediction of Raman
\cite{PE:Lev2005JCP,PE:Bar2000BOOK}, IR
\cite{PE:Yan2005JCP,PE:Man1996BOOK}, CD \cite{PE:Sre2000BOOK}, VCD
\cite{PE:Cho2005JCP,PE:Kei2000BOOK}, NMR
\cite{PE:Pel2000AB,PE:Cas1994ME} spectra, etc. are tasks that require
knowledge of the probability density in the \emph{conformational
space} only, i.e., having averaged out the external coordinates that
describe overall translations and rotations of the system.

If Cartesian coordinates are used, the integration over the momenta
produces a constant factor (which depends on the temperature $T$ but
does not depend on the coordinates) and the marginal probability
density in the coordinate space resembles the common Boltzmann
weight but using the potential energy $V(x^{\mu})$ instead of the
whole energy:

\begin{equation}
\label{eq:pc}
p_{c}(x^{\mu}) =\frac{\exp{\big[-\beta V(x^{\mu})\big]}}
                {\displaystyle \int\mathrm{d}x^{\nu}\,
                \exp{\big[-\beta V(x^{\nu})\big]}} \ .
\end{equation}

Typically the potential energy does not change under global translations
and rotations of the system. In addition, as we have already mentioned,
one is normally not interested in averages of observables that depend
on these degrees of freedom. Hence, it would be convenient to average
them out from eq.~(\ref{eq:pc}). However, this cannot be done
in Cartesian coordinates: one must use a set of coordinates adapted
to overall translations and rotations.

In the simulation of macromolecules, it is customary
\cite{PE:Ech2006JCCa,PE:Cra2002BOOK,PE:Cha2002IJQC,PE:Aba1994JCC,PE:Alm1990MP,PE:Maz1989JBSD,PE:Aba1989JBSD,PE:Go1976MM}
to define a set of curvilinear coordinates $q^{\mu}$ in which the
first six ones, denoted by $q^{A}$, are called \emph{external
coordinates} and parametrize the system overall position,
specifying the position of a selected point (normally an atom), and
rotation, via three Euler angles (see sec.~\ref{sec:definitions}). The
remaining $3n-6$ coordinates (where $n$ is the number of mass points
or \emph{atoms}) are called \emph{internal coordinates} and will be
denoted herein by $q^{a}$.

This change of coordinates modifies the mass-metric tensor in the kinetic
energy. Thus, when the momenta are averaged out and the marginal
probability density in the whole coordinate space is considered, the
square root of the determinant of the mass-metric tensor (which now does
depend on the coordinates) shows up:

\begin{equation}
\label{eq:Pw}
p_{w}(q^{\mu}) = \frac{{\det}^{\frac{1}{2}}G(q^{A},q^{a})\,
                 \exp{\big[-\beta V(q^{a})\big]}}
                 {\displaystyle \int\mathrm{d}q^{B}\mathrm{d}q^{b}\, 
                 {\det}^{\frac{1}{2}} G(q^{B},q^{b})\,
                 \exp{\big[-\beta V(q^{b})\big]}} \ .
\end{equation}

More interestingly, if holonomic constraints are imposed on the system
(the so-called \emph{classical rigid} model
\cite{PE:Ech2006JCCb,PE:Mor2004ACP,PE:Den2000MP}), the reduced
mass-metric tensor on the constrained hypersurface appears in the
kinetic energy. Hence, when the momenta are integrated out from the
joint probability density in the phase space, the square root of
its determinant occurs:

\begin{equation}
\label{eq:Pr}
p_{r}(q^{u}) = \frac{{\det}^{\frac{1}{2}}g(q^{A},q^{i})\,
               \exp{\big[-\beta V_{\Sigma}(q^{i})\big]}}
               {\displaystyle \int\mathrm{d}q^{B}\mathrm{d}q^{j}\, 
               {\det}^{\frac{1}{2}} g(q^{B},q^{j})\,
               \exp{\big[-\beta V_{\Sigma}(q^{j})\big]}} \ ,
\end{equation}

where $V_{\Sigma}$ stands for the potential energy in the constrained
hypersurface $\Sigma$, $q^{u} \equiv (q^{A},q^{i})$ denotes the
\emph{soft coordinates}, among which the external ones $q^{A}$ are
included, and $q^{i}$ denotes the soft internal coordinates.

If, on the other hand, the constraints are imposed via a steep
potential that energetically penalizes the conformations that leave
the constrained hypersurface (the so-called \emph{classical stiff}
model \cite{PE:Ech2006JCCb,PE:Mor2004ACP,PE:Den2000MP,PE:Go1976MM}),
the probability density is the same as in eq.~(\ref{eq:Pw}) except for
the determinant of the Hessian matrix $\mathcal{H}$ of the constraining
potential that appears when the \emph{hard coordinates} are averaged
out and for the fact that all the functions are evaluated on the
constrained hypersurface and, consequently, depend only on the soft
coordinates $q^{u}$:

\begin{equation}
\label{eq:Ps}
p_{s}(q^{u}) = \frac{\displaystyle
      {\det}^{\frac{1}{2}}G(q^{A},q^{i})\,
      {\det}^{-\frac{1}{2}}\mathcal{H}(q^{i})\, \exp{\big[-\beta
      V_{\Sigma}(q^{i})\big]}} {\displaystyle
      \int\mathrm{d}q^{B}\mathrm{d}q^{j}\,
      {\det}^{\frac{1}{2}}G(q^{B},q^{j})\,
      {\det}^{-\frac{1}{2}}\mathcal{H}(q^{j})\, \exp{\big[-\beta
      V_{\Sigma}(q^{j})\big]}} \ .
\end{equation}

Finally, the Fixman's compensating potential
\cite{PE:Fix1974PNAS,PE:Ech2006JCCb,PE:Mor2004ACP,PE:Den2000MP},
denoted by $V_{F}$ and which is customarily used to reproduce the
stiff equilibrium distribution using rigid molecular dynamics
simulations \cite{PE:Ech2006JCCb,PE:Alm1990MP,PE:Per1985MM}, is also
expressed as a function of these determinants:

\begin{equation}
\label{eq:VF}
V_{F}(q^{u}):=\frac{RT}{2}\ln\Bigg[
      \frac{{\det}^{\frac{1}{2}}g(q^{A},q^{i})\,
            {\det}^{\frac{1}{2}}\mathcal{H}(q^{i})}
            {{\det}^{\frac{1}{2}}G(q^{A},q^{i})} \Bigg] \ .
\end{equation}

One should note that the fields of application of the different
mass-metric tensor determinants are distinct. The Fixman's
compensating potential above is only meant to allow that the stiff
distribution in eq.~(\ref{eq:Ps}) be sampled in rigid Molecular Dynamics
simulations
\cite{PE:Pas2002JCP,PE:Chu2000JCC,PE:Rei2000PD,PE:Zho2000JCP,PE:Den2000MP,PE:He1998JCPb,PE:Per1985MM,PE:Ber1984BOOK,PE:Pea1979JCP,PE:Cha1979JCP,PE:Hel1979JCP,PE:Fix1978JCP}.
and \emph{it should never be included in Monte Carlo simulations}. On
the other hand, if one chooses as his physical description the whole
space, the rigid or the stiff model, the probability densities that
must be sampled are the ones in eqs.~(\ref{eq:Pw}), (\ref{eq:Pr}) and
(\ref{eq:Ps}), respectively
\cite{PE:Ech2006JCCb,PE:Mor2004ACP,PE:Den2000MP,PE:Ral1979JFM,PE:Hel1979JCP,PE:Go1976MM,PE:Fix1974PNAS}.
Due to the averaging out of the momenta in these expressions, the
determinants of either $G$ or $g$ show up, hence, if Monte Carlo
simulations of these models are to be performed, these corrections
(which are related to but different from the Fixman's compensating
potential) should be included or, otherwise, shown to be negligible.
The discussion found in the literature about the necessity of
including these terms \cite{PE:Pat2004JCP} is based on different
simplifications, approximations and misconceptions. On one side, since
G\=o and Scheraga showed some decades ago that the determinant of $G$,
for a serial polymer with constant bond angles and bond lengths, does
not depend on the conformation of the molecule \cite{PE:Go1976MM}, it
is customarily neglected in the literaure. However, this must be
understood as an approximation, since, as some authors have recognized
\cite{PE:Che2005JCC,PE:Hes2002JCP,PE:Zho2000JCP,PE:Go1976MM}, the
constrained values of the hard coordinates depend on the soft ones
even in the case of simple force-fields, due to the long-range energy
terms and, in such a case, $\det G$ does depend on the $q^{i}$,
rendering pertintent its inclusion in the basic equations. Moreover,
the determinant of the Hessian $\mathcal{H}$ of the constraining part
of the potential is also assumed to be independent of conformation by
most authors \cite{PE:Den2000MP,PE:Ber1983BOOK,PE:Go1976MM}. This is
also an approximation, due to the same reasons presented above, and
should be assessed in each case, however, if these two approximations
are made (the neglect of $\det G$ and of $\det \mathcal{H}$), the
Fixman's compensating potential in eq.~(\ref{eq:VF}) depends only on
$\det g$, just as the correcting term in the rigid case. This is one
of the sources of confusion and the reason that the relevance of
mass-metric effects is associated only with the conformational
dependence of the determinant of the reduced mass-metric tensor $g$
\cite{PE:Pat2004JCP}. One should also note that most authors accept
that the \emph{correct} constrained model is the stiff one
\cite{PE:Pat2004JCP,PE:Rei2000PD,PE:Den2000MP,PE:Ber1983BOOK,PE:VGu1982MM,PE:Pea1979JCP,PE:Fix1974PNAS},
and, therefore, no works are written in which the rigid probability
density in eq.~(\ref{eq:Pr}) is sampled. In our opinion, the question
whether the rigid or the stiff model should be used to approximate the
real quantum mechanical statistics of an arbitrary organic molecule
has not been satisfactorily answered yet. For discussions about the
topic, see references
\cite{PE:Mor2004ACP,PE:Pea1979JCP,PE:Pea1979JCP,PE:Hes2002JCP,PE:Ral1979JFM,PE:Hel1979JCP,PE:Go1976MM,PE:Go1969JCP}.
In this work, we adopt the cautious position that any of the two
models may be useful in certain cases or for certain purposes and we
study them both on equal footing. All these approximations together
are the reason of the fact that no determinants are included in Monte
Carlo simulations and that the Fixman's compensating potential is
erroneusly regarded as playing a role outside Molecular Dynamics
simulations.

In the three physical models, given by eqs.~(\ref{eq:Pw}),
(\ref{eq:Pr}) and (\ref{eq:Ps}), and in the Fixman's potential written
above, neither the potential energy nor the Hessian matrix of the
constraining potential depend on the external coordinates. Therefore,
it would be very convenient to integrate them out in order to obtain a
simpler probability density depending only on the internal
coordinates. Such an improvement may render the coding of computer
applications and the visualization of molecules easier, since all the
movements in the molecule may be performed fixing an atom in space and
keeping the orientation of the system with respect to a set of axes
fixed in space constant \cite{PE:Wil1980BOOK}. Moreover, the insight
gained and the usefulness of the calculations herein in further
analytical studies constitute additional benefits. Regarding Monte
Carlo simulations, it is not clear that much computational effort will
be saved for macromolecules, since the gain expected when reducing the
degrees of freedom from $M+6$ to $M$ (where $M$ is the number of soft
internal coordinates) is only appreciable if $M$ is small and the
difficulties arising from the use of curvilinear coordinates may well
be more important. However, in cases where the use of curvilinear
coordinates is a must, such as the simulation of constrained systems,
the calculations in this work allow to save a time (which will depend
on the size of the system) that, otherwise, would be wasted in
movements of the external coordinates. Finally, if the molecule
treated is small, as it is common in ab initio Quantum Mechanical
calculations in model peptides
\cite{PE:Lan2005PSFB,PE:Per2003JCC,PE:Var2002JPCA,PE:Yu2001JMS,PE:Csa1999PBMB},
the relevance of omitting the externals may be considerable. In a work
recently done in our group \cite{PE:Ech2006JCCb}, in which the model
dipeptide HCO-{\small L}-Ala-NH$_{2}$ is studied, the soft internals
are two: the Ramachandran angles $\phi$ and $\psi$; hence, $M=2$ and
the formulae in this work have permitted to reduce the number of
degrees of freedom from 8 to 2. This situation is very common in the
literature
\cite{PE:Lan2005PSFB,PE:Per2003JCC,PE:Var2002JPCA,PE:Yu2001JMS,PE:Csa1999PBMB}.

Now, for the resulting expressions to be manageable, the determinant
of the mass-metric tensor $G$, in $p_{w}$ or $p_{s}$, and the
determinant of $g$, in the rigid case, should \emph{factorize} as a
product of a function that depends only on the external coordinates
and another function that depends only on the internal ones. Then the
function depending on the external coordinates, could be integrated
out in the probability densities $p_{w}$, $p_{r}$ and $p_{s}$ or taken
out of the logarithm in $V_{F}$\footnote{What really happens, (see
secs.~\ref{sec:constrained} and \ref{sec:unconstrained}) is that the
factor that depends on the external coordinates is the same for $\det
G$ and $\det g$. Hence, it divides out in eq.~(\ref{eq:VF}) (see
sec.~\ref{sec:conclusions}).}.

For some simple examples, it has already been proven in the literature
that this factorization actually happens. In
ref.~\citen{PE:Go1976MM}, the determinant of $G$ is shown to
factorize for a serial polymer in a particular set of curvilinear
coordinates. In ref.~\citen{PE:Pat2004JCP}, the determinant of
$g$ is shown to factorize for the same system, in similar coordinates,
with frozen bond lengths and bond angles.

In this work, we \emph{generalize} these results, showing that they
hold in \emph{arbitrary} internal coordinates (for general branched
molecules) and with \emph{arbitrary} constraints. Perhaps more
importantly, we provide explicit expression for the functions involved
in the factorization. It is worth remarking that, although the
calculations herein have been performed thinking in macromolecules as
target system, they are completely general and applicable to any
classical system composed by discrete mass points.

In sec.~\ref{sec:definitions}, we present the notation and conventions
that will be used throughout the article. In
sec.~\ref{sec:constrained}, we explicitly factorize the determinant of
the reduced mass-metric tensor $g$ as a product of a function that
depends only on the external coordinates and another function that
depends on arbitrary internal coordinates; no specific form is assumed
for the constraints. In sec.~\ref{sec:unconstrained}, we perform the
analogous calculations for the determinant of the mass-metric tensor
$G$ and sec.~\ref{sec:conclusions} is devoted to the conclusions. In
appendix A, the general mathematical argument underlying these results
is given. Finally, in appendix B, the determinant of the mass-metric
tensor $G$ is computed in the set of curvilinear coordinates
introduced in ref.~\citen{PE:Ech2006JCCa}, which turn out to be
convenient for dealing with general branched molecules. Moreover, we
show that the classical formula for serial polymers \cite{PE:Go1976MM}
is actually valid for any macromolecule.

\section{General set-up and definitions}
\label{sec:definitions}

The system under scrutiny will be a set of $n$ mass points termed
\emph{atoms}. This section is devoted to introduce certain notational
conventions that will be used extensively in the rest of the paper.

\begin{itemize}

\item The superindex $T$ indicates matrix transposition. By
$\vec{a}^{\,T}$ we shall understand the row vector
$(a^{1},a^{2},a^{3})$.

\item The Cartesian coordinates of the atom $\alpha$ in a set of axes
fixed in space are denoted by $\vec{x}_\alpha$. The subscript $\alpha$
runs from 1 to $n$.

\item The curvilinear coordinates suitable to integrate out the
external degrees of freedom will be denoted by $q^{\mu},\
\mu=1,\ldots,3n$. We shall often use $N:=3n$ for the total number of
degrees of freedom.

\item We choose the coordinates $q^{\mu}$ so that the first six are
\emph{external coordinates}. They are denoted by $q^{A}$ and their
ordering is $q^{A} \equiv (X,Y,Z,\phi,\theta,\psi)$. The first three
ones, $\vec{X}^{T}:=(X,Y,Z)$, describe the overall position of the
system. The three angles $(\phi,\theta,\psi)$ are related to its
overall orientation. More concretely, they give the orientation of a
frame fixed in the system with respect to the frame fixed in space.

\item To define the set of axes \emph{fixed in the system}, we select
three atoms (denoted by 1, 2 and 3) in such a way that $\vec{X}$ is
the position of atom 1 (i.e., $\vec{x}_{1} = \vec{X}$). The
orientation of the fixed axes
$(x^{\,\prime},y^{\,\prime},z^{\,\prime})$ is chosen such that atom
2 lies in the positive half of the $z^{\,\prime}$-axis and atom 3 is
contained in the $(x^{\,\prime},z^{\,\prime})$-plane, in the positive
half of the $x^{\,\prime}$-axis (see fig.~\ref{fig:axes}). The
position of atom $\alpha$ in these axes is denoted by
$\vec{x}_{\alpha}^{\,\prime}$.

\begin{figure}[!ht]
\begin{center}
\epsfig{file=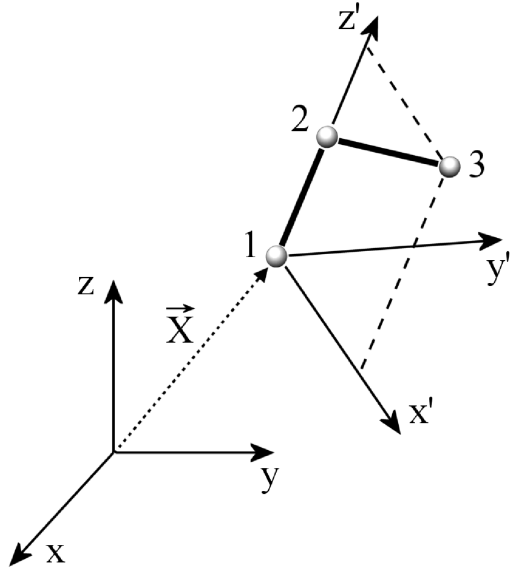,width=7cm}
\end{center}
\caption{\label{fig:axes}{\small Definition of the axes fixed in the system.}}
\end{figure}

\item Let $E(\phi,\theta,\psi)$ be the Euler rotation matrix (in the
ZYZ convention) that takes a vector of primed components
$\vec{a}^{\,\prime}$ to the frame fixed in space, i.e.,
$\vec{a}=E(\phi,\theta,\psi)\,\vec{a}^{\,\prime}$. Its explicit
expression is the following:

\begin{equation}
\label{eq:def_E}
E(\phi,\theta,\psi)=
 \underbrace{
 \left( \begin{smallmatrix}
 \scriptstyle \cos\phi & -\sin\phi& 0 \\
 \sin\phi & \cos\phi & 0 \\
 0 & 0 & 1
 \end{smallmatrix} \right)
 }_{\displaystyle \Phi(\phi)}
 \underbrace{
 \left( \begin{smallmatrix}
 -\cos\theta & 0 & \sin\theta \\
 0 & 1 & 0 \\
 -\sin\theta & 0 & -\cos\theta
 \end{smallmatrix} \right)
 }_{\displaystyle \Theta(\theta)}
 \underbrace{
 \left( \begin{smallmatrix}
 \cos\psi & -\sin\psi& 0 \\
 \sin\psi & \cos\psi & 0 \\
 0 & 0 & 1
 \end{smallmatrix} \right)
 }_{\displaystyle \Psi(\psi)}
\ .
\end{equation}

The unusual minus signs of the cosines in the diagonal of matrix
$\Theta(\theta)$ come from the fact that, due to frequent biochemical
conventions, the rotation with respect to the $y$-axis is of angle
$\tilde{\theta}:=\pi - \theta$.

\item The coordinates $q^{\mu}$ are split into $(q^{A},q^{a}),\
a=7,\dots,N$. The coordinates $q^{a}$ are said {\it internal
coordinates} and determine the positions of the atoms in the frame
fixed in the system. The transformation from the Cartesian coordinates
$\vec{x}_\alpha$ to the curvilinear coordinates $q^{\mu}$ may be written
as follows:

\begin{equation}
\label{eq:transf1_unconstrained}
\vec{x}_{\alpha} = \vec{X} + E(\phi,\theta,\psi)\,
 \vec{x}^{\,\prime}_{\alpha}(q^{a}) \qquad
 \alpha = 1,\ldots,n \ .
\end{equation}

\item The coordinates $q^{a}$ parameterize what we shall call the {\it
internal subspace}, denoted by $\cal I$. Assume that $L$ independent
constraints are imposed on $\cal I$, so that only points on a
hypersurface $\Sigma\subset{\cal I}$ of dimension $M:=N-L-6$ are
allowed. Then, we choose a splitting $q^{a} \equiv (q^{i},q^{I})$, with
$i=7,\dots,M+6$ and $I=M+7,\dots,N$, where $q^{i}$ (the {\it internal
soft coordinates}) parameterize $\Sigma$, and $q^{I}$ (the {\it hard
coordinates}) are functions of the soft coordinates:

\begin{equation}
\label{eq:constraints}
q^{I}=f^{I}(q^{i}) \qquad I=M+7,\ldots,N \ .
\end{equation}

If these constraints are used, together with
eq.~(\ref{eq:transf1_unconstrained}), the Cartesian position of any atom
may be parameterized with the set of \emph{all soft coordinates},
denoted by $q^{u} \equiv (q^{A},q^{i})$, with $u=1,\ldots,M+6$, as
follows:

\begin{equation}
\label{eq:transf1_constrained}
\vec{x}_{\alpha} = \vec{X} + E(\phi,\theta,\psi)\,
\vec{x}^{\,\prime}_{\alpha}\big(q^{i},f^{I}(q^{i})\big) \qquad
 \alpha = 1,\ldots,n \ .
\end{equation}

\item In table~\ref{tab:def_indices}, a summary of the indices
 used is given.

\end{itemize}

\begin{table}[!ht]
\begin{center}
\begin{tabular}{llll}
Indices & Range & Number & Description \\
\hline \\[-8pt]
$\alpha,\beta,\gamma,\ldots$ & $1,\ldots,n$ & $n$ &Atoms \\
$p,q,r,s,\ldots$ & $1,2,3$ & 3 & Components of trivectors \\
$\mu,\nu,\rho,\ldots$ & $1,\ldots,N$ & $N=3n$ & All coordinates \\
$A,B,C,\ldots$ & $1,\ldots,6$ & 6 & External coordinates \\
$a,b,c,\ldots$ & $7,\ldots,N$ & $N-6$ & Internal coordinates \\
$i,j,k,\ldots$ & $7,\ldots,M+6$ & $M$ & Soft internal coordinates \\
$I,J,K,\ldots$ & $M+7,\ldots,N$ & $L=N-M-6$ & Hard internal coordinates \\
$u,v,w,\ldots$ & $1,\ldots,M+6$ & $M+6$ & All soft coordinates
\end{tabular}
\end{center}
\caption{\label{tab:def_indices}{\small Definition of the indices used.}}
\end{table}

\section{Constrained case}
\label{sec:constrained}

The \emph{reduced mass-metric tensor}, in the constrained hypersurface
$\Sigma$ plus the external subspace spanned by the $q^{A}$, may be
written as follows:

\begin{equation}
\label{eq:reduced_metric}
g_{vw}(q^{u}):=\sum_{\mu=1}^{N} 
 \frac{\partial x^{\mu}(q^{u})}{\partial q^{v}}
 m_{\mu}\frac{\partial x^{\mu}(q^{u})}{\partial q^{w}} \ .
\end{equation}

In matrix notation, this is written as

\begin{equation}
\label{eq:reduced_metric_matrix}
g=J_{c}^{T}MJ_{c} \ ,
\end{equation}

where $c$ stands for \emph{constrained} and $M$ is the diagonal $N
\times N$ \emph{mass matrix} given by

\begin{equation}
\label{eq:mass_matrix}
M:=\left(
\begin{array}{ccc}
m_{1}^{(3)} & & 0 \\
& \ddots & \\
0 & & m_{n}^{(3)}
\end{array}
\right)
\ , \quad \mathrm{with} \quad
m_{\alpha}^{(3)}:=m_{\alpha}
\underbrace{\left(
\begin{array}{ccc}
1 & 0 & 0 \\
0 & 1 & 0 \\
0 & 0 & 1
\end{array}
\right)}_{\displaystyle I^{(3)}} \ .
\end{equation}

Using eq.~(\ref{eq:transf1_constrained}) and noting that the
derivatives with respect to the externals, $q^{A}$, only affect the
$\vec{X}$ vector and the Euler rotation matrix $E$, while
differentiation with respect to soft internals, $q^{i}$, only act
on the $\vec{x}^{\,\prime}_{\alpha}$, we have that $J_{c}$ is the
$N \times (M+6)$ matrix

\begin{equation}
\label{eq:Jc}
J_{c}=\left(
\begin{array}{cccc|c@{\hspace{2pt}}c@{\hspace{2pt}}c}
I^{(3)} & & 0 & & & 0 & \\[5pt]
I^{(3)} & \displaystyle \frac{\partial E}{\partial \phi}\vec{x}^{\,\prime}_{2}
& \displaystyle \frac{\partial E}{\partial \theta}\vec{x}^{\,\prime}_{2}
& \displaystyle \frac{\partial E}{\partial \psi}\vec{x}^{\,\prime}_{2}
& \cdots &
\displaystyle E\frac{\partial \vec{x}^{\,\prime}_{2}}{\partial q^{j}} & 
\cdots \\[-2pt]
\vdots & \vdots & \vdots & \vdots & & \vdots & \\
I^{(3)} & \displaystyle \frac{\partial E}{\partial \phi}
 \vec{x}^{\,\prime}_{\alpha}
& \displaystyle \frac{\partial E}{\partial \theta}\vec{x}^{\,\prime}_{\alpha}
& \displaystyle \frac{\partial E}{\partial \psi}\vec{x}^{\,\prime}_{\alpha}
& \cdots &
\displaystyle E\frac{\partial \vec{x}^{\,\prime}_{\alpha}}{\partial q^{j}} & 
\cdots \\[-2pt]
\vdots & \vdots & \vdots & \vdots & & \vdots &
\end{array}
\right)
\ .
\end{equation}

If we perform the matrix multiplications in
eq.~(\ref{eq:reduced_metric_matrix}), we obtain

\begin{equation}
\label{eq:g}
g=\!\left(
\begin{array}{cc|c@{\hspace{2pt}}c@{\hspace{2pt}}c}
\displaystyle I^{(3)}\mathop{\Sigma}_{\alpha}m_{\alpha} &
\displaystyle \mathop{\Sigma}_{\alpha}m_{\alpha}\partial E_{\alpha} &
\cdots & \displaystyle \mathop{\Sigma}_{\alpha}m_{\alpha}E
\frac{\partial \vec{x}^{\,\prime}_{\alpha}}{\partial q^{j}}& \cdots \\[10pt]
\displaystyle \mathop{\Sigma}_{\alpha}m_{\alpha}\partial E_{\alpha}^{T} &
\displaystyle \mathop{\Sigma}_{\alpha}m_{\alpha}
\partial E_{\alpha}^{T}\partial E_{\alpha} &
\cdots & \displaystyle \mathop{\Sigma}_{\alpha}m_{\alpha}
\partial E_{\alpha}^{T}E
\frac{\partial \vec{x}^{\,\prime}_{\alpha}}{\partial q^{j}} & \cdots \\[10pt]
\hline
\vdots & \vdots & & \vdots & \\
\displaystyle \mathop{\Sigma}_{\alpha}m_{\alpha}
\frac{\partial \vec{x}^{\,\prime\,T}_{\alpha}}{\partial q^{i}}E^{T} &
\displaystyle \mathop{\Sigma}_{\alpha}m_{\alpha}
\frac{\partial \vec{x}^{\,\prime\,T}_{\alpha}}{\partial q^{i}}E^{T}
\partial E_{\alpha} & \cdots &
\displaystyle \mathop{\Sigma}_{\alpha}m_{\alpha}
\frac{\partial \vec{x}^{\,\prime\,T}_{\alpha}}{\partial q^{i}}
\frac{\partial \vec{x}^{\,\prime}_{\alpha}}{\partial q^{j}} & \cdots \\[-2pt]
\vdots & \vdots & & \vdots & 
\end{array}
\right)
\ .
\end{equation}

All the sums in $\alpha$ can be understood as ranging from 1 to $n$ if
we note that $\vec{x}^{\,\prime}_{1}=\vec{0}$.  In the bottom right
block, the fact that $E$ is an orthogonal matrix (i.e., that
$E^{T}E=I^{(3)}$) has been used, and we have defined the $3 \times 3$
block as

\begin{equation}
\label{eq:dE}
\partial E_{\alpha}:=\left(
\frac{\partial E}{\partial \phi}\vec{x}^{\,\prime}_{\alpha}\ \
\frac{\partial E}{\partial \theta}\vec{x}^{\,\prime}_{\alpha}\ \
\frac{\partial E}{\partial \psi}\vec{x}^{\,\prime}_{\alpha}\right) \ .
\end{equation}

We can write $g$ as

\begin{equation}
\label{eq:Eg1E}
g:=
\left( \begin{array}{cc}
E & 0 \\
0 & I^{(M+3)}
\end{array} \right)
g_{1}
\left( \begin{array}{cc}
E^{T} & 0 \\
0 & I^{(M+3)}
\end{array} \right)
\ ,
\end{equation}

where $I^{(M+3)}$ is the $(M+3) \times (M+3)$ identity matrix and
$g_{1}$ is defined as

\begin{equation}
\label{eq:g1}
g_{1}:=\left(
\begin{array}{cc|c@{\hspace{2pt}}c@{\hspace{2pt}}c}
\displaystyle I^{(3)}\mathop{\Sigma}_{\alpha}m_{\alpha} &
\displaystyle \mathop{\Sigma}_{\alpha}m_{\alpha}E^{T}\partial E_{\alpha} &
\cdots & \displaystyle \mathop{\Sigma}_{\alpha}m_{\alpha}
\frac{\partial \vec{x}^{\,\prime}_{\alpha}}{\partial q^{j}}& \cdots \\[10pt]
\displaystyle \mathop{\Sigma}_{\alpha}m_{\alpha}\partial E_{\alpha}^{T}E &
\displaystyle \mathop{\Sigma}_{\alpha}m_{\alpha}
\partial E_{\alpha}^{T}EE^{T}\partial E_{\alpha} &
\cdots & \displaystyle \mathop{\Sigma}_{\alpha}m_{\alpha}
\partial E_{\alpha}^{T}E
\frac{\partial \vec{x}^{\,\prime}_{\alpha}}{\partial q^{j}} & \cdots \\[10pt]
\hline
\vdots & \vdots & & \vdots & \\
\displaystyle \mathop{\Sigma}_{\alpha}m_{\alpha}
\frac{\partial \vec{x}^{\,\prime\,T}_{\alpha}}{\partial q^{i}} &
\displaystyle \mathop{\Sigma}_{\alpha}m_{\alpha}
\frac{\partial \vec{x}^{\,\prime\,T}_{\alpha}}{\partial q^{i}}E^{T}
\partial E_{\alpha} & \cdots &
\displaystyle \mathop{\Sigma}_{\alpha}m_{\alpha}
\frac{\partial \vec{x}^{\,\prime\,T}_{\alpha}}{\partial q^{i}}
\frac{\partial \vec{x}^{\,\prime}_{\alpha}}{\partial q^{j}} & \cdots \\[-2pt]
\vdots & \vdots & & \vdots & 
\end{array}
\right)
\ .
\end{equation}

Note that $I^{(3)}=EE^{T}$ has been introduced in the bottom right
$3 \times 3$ submatrix of the top left block.

Next, we introduce some simplifying notation for the
matrices $E^{T}\partial E_{\alpha}$:

\begin{equation}
\label{eq:ETdE}
E^{T}\partial E_{\alpha}=\biggl(
E^{T}\frac{\partial E}{\partial \phi}\vec{x}^{\,\prime}_{\alpha}\ \
E^{T}\frac{\partial E}{\partial \theta}\vec{x}^{\,\prime}_{\alpha}\ \
E^{T}\frac{\partial E}{\partial \psi}\vec{x}^{\,\prime}_{\alpha}\biggr)=:
\biggl( \vec{y}^{\,1}_{\alpha}\ \vec{y}^{\,2}_{\alpha}\
       \vec{y}^{\,3}_{\alpha} \biggr) \ .
\end{equation}

Defining

\begin{equation}
\label{eq:g1alpha}
g_{1}^{\alpha}:=\left(
\begin{array}{cccc|c@{\hspace{2pt}}c@{\hspace{2pt}}c}
\displaystyle I^{(3)} &
\displaystyle \vec{y}^{\,1}_{\alpha} &
\displaystyle \vec{y}^{\,2}_{\alpha} &
\displaystyle \vec{y}^{\,3}_{\alpha} &
\cdots & \displaystyle 
\frac{\partial \vec{x}^{\,\prime}_{\alpha}}{\partial q^{j}}& \cdots \\[10pt]
\displaystyle \vec{y}^{\,1\,T}_{\alpha} &
\displaystyle \vec{y}^{\,1\,T}_{\alpha}\vec{y}^{\,1}_{\alpha} &
\displaystyle \vec{y}^{\,1\,T}_{\alpha}\vec{y}^{\,2}_{\alpha} &
\displaystyle \vec{y}^{\,1\,T}_{\alpha}\vec{y}^{\,3}_{\alpha} &
\cdots & \displaystyle \vec{y}^{\,1\,T}_{\alpha}
\frac{\partial \vec{x}^{\,\prime}_{\alpha}}{\partial q^{j}} & \cdots \\[10pt]
\displaystyle \vec{y}^{\,2\,T}_{\alpha} &
\displaystyle \vec{y}^{\,2\,T}_{\alpha}\vec{y}^{\,1}_{\alpha} &
\displaystyle \vec{y}^{\,2\,T}_{\alpha}\vec{y}^{\,2}_{\alpha} &
\displaystyle \vec{y}^{\,2\,T}_{\alpha}\vec{y}^{\,3}_{\alpha} &
\cdots & \displaystyle \vec{y}^{\,2\,T}_{\alpha}
\frac{\partial \vec{x}^{\,\prime}_{\alpha}}{\partial q^{j}} & \cdots \\[10pt]
\displaystyle \vec{y}^{\,3\,T}_{\alpha} &
\displaystyle \vec{y}^{\,3\,T}_{\alpha}\vec{y}^{\,1}_{\alpha} &
\displaystyle \vec{y}^{\,3\,T}_{\alpha}\vec{y}^{\,2}_{\alpha} &
\displaystyle \vec{y}^{\,3\,T}_{\alpha}\vec{y}^{\,3}_{\alpha} &
\cdots & \displaystyle \vec{y}^{\,3\,T}_{\alpha}
\frac{\partial \vec{x}^{\,\prime}_{\alpha}}{\partial q^{j}} & \cdots \\[10pt]
\hline
\vdots & \vdots & \vdots & \vdots & & \vdots & \\
\displaystyle 
\frac{\partial \vec{x}^{\,\prime\,T}_{\alpha}}{\partial q^{i}} &
\displaystyle \frac{\partial \vec{x}^{\,\prime\,T}_{\alpha}}{\partial q^{i}}
\vec{y}^{\,1}_{\alpha} &
\displaystyle \frac{\partial \vec{x}^{\,\prime\,T}_{\alpha}}{\partial q^{i}}
\vec{y}^{\,2}_{\alpha} &
\displaystyle \frac{\partial \vec{x}^{\,\prime\,T}_{\alpha}}{\partial q^{i}}
\vec{y}^{\,3}_{\alpha} &
\cdots &
\displaystyle
\frac{\partial \vec{x}^{\,\prime\,T}_{\alpha}}{\partial q^{i}}
\frac{\partial \vec{x}^{\,\prime}_{\alpha}}{\partial q^{j}} & \cdots \\[-2pt]
\vdots & \vdots & \vdots & \vdots & & \vdots & 
\end{array}
\right)
\ ,
\end{equation}

we have that

\begin{equation}
\label{eq:g1sumg1alpha}
g_{1}=\sum_{\alpha} m_{\alpha} g_{1}^{\alpha} \ .
\end{equation}

Now, the vectors $\vec{y}^{\,p}_{\alpha}$ may be extracted
from $g_{1}^{\alpha}$ as follows:

\begin{equation}
\label{eq:YmatrixY}
g_{1}^{\alpha}=Y^{T}_{\alpha}\left(
\begin{array}{cc|c@{\hspace{2pt}}c@{\hspace{2pt}}c}
\displaystyle I^{(3)} & \displaystyle I^{(3)} &
\cdots & \displaystyle
\frac{\partial \vec{x}^{\,\prime}_{\alpha}}{\partial q^{j}} & \cdots \\[10pt]
\displaystyle I^{(3)} & \displaystyle I^{(3)} &
\cdots & \displaystyle
\frac{\partial \vec{x}^{\,\prime}_{\alpha}}{\partial q^{j}} & \cdots \\[10pt]
\hline
\vdots & \vdots & & \vdots & \\
\displaystyle
\frac{\partial \vec{x}^{\,\prime\,T}_{\alpha}}{\partial q^{i}} &
\displaystyle
\frac{\partial \vec{x}^{\,\prime\,T}_{\alpha}}{\partial q^{i}} &
\cdots &
\displaystyle
\frac{\partial \vec{x}^{\,\prime\,T}_{\alpha}}{\partial q^{i}}
\frac{\partial \vec{x}^{\,\prime}_{\alpha}}{\partial q^{j}} & \cdots \\[-2pt]
\vdots & \vdots & & \vdots & 
\end{array}
\right)Y_{\alpha}
\ ,
\end{equation}

where

\begin{equation}
\label{eq:Y}
Y_{\alpha}:=
\left( \begin{array}{cc@{\hspace{2pt}}c@{\hspace{2pt}}cc}
I^{(3)} & & 0 & & 0 \\
0 & \vec{y}^{\,1}_{\alpha} & \vec{y}^{\,2}_{\alpha} &
\vec{y}^{\,3}_{\alpha} & 0 \\
0 & & 0 & & I^{(M)}
\end{array} \right)
\ ,
\end{equation}

and the central matrix in eq.~(\ref{eq:YmatrixY}) only depends on the
soft internal coordinates.

After some lengthy calculations, one shows that

\begin{equation}
\label{eq:YvW}
\biggl( \vec{y}^{\,1}_{\alpha}\ \vec{y}^{\,2}_{\alpha}\
       \vec{y}^{\,3}_{\alpha} \biggr) =
\underbrace{
\left( \begin{array}{c@{\hspace{4pt}}c@{\hspace{4pt}}c}
0 & -x^{\,\prime\,3}_{\alpha} & x^{\,\prime\,2}_{\alpha} \\
x^{\,\prime\,3}_{\alpha} & 0 & -x^{\,\prime\,1}_{\alpha} \\
-x^{\,\prime\,2}_{\alpha} & x^{\,\prime\,1}_{\alpha} & 0
\end{array} \right)}_{\displaystyle v(\vec{x}^{\,\prime}_{\alpha})}
\underbrace{
\left( \begin{array}{c@{\hspace{6pt}}c@{\hspace{6pt}}c}
\sin\theta\cos\psi & \sin\psi & 0 \\
-\sin\theta\sin\psi & \cos\psi & 0 \\
\cos\theta & 0 & -1
\end{array} \right)}_{\displaystyle W(\theta,\psi)}
\ .
\end{equation}

Thus, the matrix $Y_{\alpha}$ in eq.~(\ref{eq:Y}) may be written as

\begin{equation}
\label{eq:fullYvW}
Y_{\alpha}=
\left( \begin{array}{ccc}
I^{(3)} & 0 & 0 \\
0 & v(\vec{x}^{\,\prime}_{\alpha}) & 0 \\
0 & 0 & I^{(M)}
\end{array} \right)
\left( \begin{array}{ccc}
I^{(3)} & 0 & 0 \\
0 & W(\theta,\psi) & 0 \\
0 & 0 & I^{(M)}
\end{array} \right)
\ .
\end{equation}

If we take this expression to eq.~(\ref{eq:YmatrixY}) and
use that, for any pair of vectors $\vec{a}$ and $\vec{b}$,
$\vec{a}^{\,T}v\bigl(\vec{b}\,\bigr) = \bigl(\vec{a} \times
\vec{b}\,\bigr)^{T}$ and $v^{T}\bigl(\vec{b}\,\bigr)\,\vec{a} =
\vec{a} \times \vec{b}$, where $\times$ denotes the usual vector cross
product, we may rewrite eq.~(\ref{eq:YmatrixY}) as follows:

\begin{equation}
\label{eq:Wg2alphaW}
g_{1}^{\alpha}=
\left( \begin{array}{ccc}
I^{(3)} & 0 & 0 \\
0 & W^{T}(\theta,\psi) & 0 \\
0 & 0 & I^{(M)}
\end{array} \right)
g_{2}^{\alpha}
\left( \begin{array}{ccc}
I^{(3)} & 0 & 0 \\
0 & W(\theta,\psi) & 0 \\
0 & 0 & I^{(M)}
\end{array} \right)
\ ,
\end{equation}

where $g_{2}^{\alpha}$ is defined as

\begin{equation}
\label{eq:g2alpha}
g_{2}^{\alpha}=
\left( \begin{array}{cc|c@{\hspace{2pt}}c@{\hspace{2pt}}c}
\displaystyle I^{(3)} & \displaystyle v(\vec{x}^{\,\prime}_{\alpha}) &
\cdots & \displaystyle
\frac{\partial \vec{x}^{\,\prime}_{\alpha}}{\partial q^{j}} & \cdots \\[10pt]
\displaystyle v^{T}(\vec{x}^{\,\prime}_{\alpha}) &
\displaystyle v^{T}(\vec{x}^{\,\prime}_{\alpha})
              v(\vec{x}^{\,\prime}_{\alpha}) &
\cdots & \displaystyle
\frac{\partial \vec{x}^{\,\prime}_{\alpha}}{\partial q^{j}} \times
\vec{x}^{\,\prime}_{\alpha} & \cdots \\[10pt]
\hline
\vdots & \vdots & & \vdots & \\
\displaystyle
\frac{\partial \vec{x}^{\,\prime\,T}_{\alpha}}{\partial q^{i}} &
\displaystyle \left(
\frac{\partial \vec{x}^{\,\prime}_{\alpha}}{\partial q^{i}} \times
\vec{x}^{\,\prime}_{\alpha} \right)^{T}&
\cdots &
\displaystyle
\frac{\partial \vec{x}^{\,\prime\,T}_{\alpha}}{\partial q^{i}}
\frac{\partial \vec{x}^{\,\prime}_{\alpha}}{\partial q^{j}} & \cdots \\[-2pt]
\vdots & \vdots & & \vdots & 
\end{array} \right) \ .
\end{equation}

At this point, we insert eq.~(\ref{eq:Wg2alphaW}) in
eq.~(\ref{eq:g1sumg1alpha}) and take the $W$ matrices out of the
sum. Since $\det W(\theta,\psi) = \det W^{T}(\theta,\psi) =
-\sin\theta$, we obtain

\begin{equation}
\label{eq:detg1}
\det g_{1} = \sin^{2}\theta \, \det \underbrace{\left(\sum_{\alpha}m_{\alpha}
 g_{2}^{\alpha}\right)}_{\displaystyle g_{2}} \ .
\end{equation}

Recalling that $\det E = \det E^{T} = 1$, from
eq.~(\ref{eq:Eg1E}), we have that

\begin{equation}
\label{eq:detg}
\det g(q^{A},q^{i}) = \det g_{1}(q^{A},q^{i}) = \sin^{2}\theta \, 
 \det g_{2}(q^{i}) \ ,
\end{equation}

and the factorization of the external coordinates has been
finally accomplished, since $g_{2}$ is the following matrix, which
depends only on the soft internal coordinates $q^{i}$:

\begin{equation}
\label{eq:g2}
g_{2}=
\left( \begin{array}{cc|c@{\hspace{2pt}}c@{\hspace{2pt}}c}
m_{tot}\displaystyle I^{(3)} & \displaystyle m_{tot}\,v(\vec{R}) &
\cdots & \displaystyle m_{tot} \frac{\partial \vec{R}}{\partial q^{j}} &
\cdots \\[10pt]
\displaystyle m_{tot}\,v^{T}(\vec{R}) & \mathcal{J} &
\cdots & \displaystyle \mathop{\Sigma}_{\alpha}m_{\alpha}
\frac{\partial \vec{x}^{\,\prime}_{\alpha}}{\partial q^{j}} \times
\vec{x}^{\,\prime}_{\alpha} & \cdots \\[10pt]
\hline
\vdots & \vdots & & \vdots & \\
\displaystyle m_{tot}\frac{\partial \vec{R}}{\partial q^{i}} &
\displaystyle \mathop{\Sigma}_{\alpha}m_{\alpha} \left(
\frac{\partial \vec{x}^{\,\prime}_{\alpha}}{\partial q^{i}} \times
\vec{x}^{\,\prime}_{\alpha} \right)^{T}&
\cdots &
\displaystyle \mathop{\Sigma}_{\alpha}m_{\alpha}
\frac{\partial \vec{x}^{\,\prime\,T}_{\alpha}}{\partial q^{i}}
\frac{\partial \vec{x}^{\,\prime}_{\alpha}}{\partial q^{j}} & \cdots \\[-2pt]
\vdots & \vdots & & \vdots & 
\end{array} \right) \ ,
\end{equation}

where we have defined the \emph{total mass} of the system $m_{tot} :=
\sum_{\alpha}m_{\alpha}$, the position of the \emph{center of mass} of
the system in the primed reference frame $\vec{R} :=
m^{-1}_{tot}\sum_{\alpha}m_{\alpha} \vec{x}^{\,\prime}_{\alpha}$
and the \emph{inertia tensor} of the system, also in the primed
reference frame:

\begin{equation}
\label{eq:inertia}
\mathcal{J}\!:=\!\!
\left( \begin{smallmatrix}
\sum_{\alpha}m_{\alpha} ((x^{\,\prime\,2}_{\alpha})^{2}+
                        (x^{\,\prime\,3}_{\alpha})^{2}) &
-\sum_{\alpha}m_{\alpha}x^{\,\prime\,1}_{\alpha}
                        x^{\,\prime\,2}_{\alpha} &
-\sum_{\alpha}m_{\alpha}x^{\,\prime\,1}_{\alpha}
                        x^{\,\prime\,3}_{\alpha} \\
-\sum_{\alpha}m_{\alpha}x^{\,\prime\,1}_{\alpha}
                        x^{\,\prime\,2}_{\alpha} &
\sum_{\alpha}m_{\alpha} ((x^{\,\prime\,1}_{\alpha})^{2}+
                        (x^{\,\prime\,3}_{\alpha})^{2}) &
-\sum_{\alpha}m_{\alpha}x^{\,\prime\,2}_{\alpha}
                        x^{\,\prime\,3}_{\alpha} \\
-\sum_{\alpha}m_{\alpha}x^{\,\prime\,1}_{\alpha}
                        x^{\,\prime\,3}_{\alpha} &
-\sum_{\alpha}m_{\alpha}x^{\,\prime\,2}_{\alpha}
                        x^{\,\prime\,3}_{\alpha} &
\sum_{\alpha}m_{\alpha} ((x^{\,\prime\,1}_{\alpha})^{2}+
                        (x^{\,\prime\,2}_{\alpha})^{2}) \\
\end{smallmatrix} \right)
 .
\end{equation}

\section{Unconstrained case}
\label{sec:unconstrained}

If no constraints are assumed and the system lives in the whole
internal space $\cal I$ plus the external subspace spanned by the
$q^{A}$, the Cartesian coordinates of the $n$ atoms must be expressed
using eq.~(\ref{eq:transf1_unconstrained}), instead of
eq.~(\ref{eq:transf1_constrained}).

We now wish to calculate the determinant of the \emph{whole-space}
mass-metric tensor in the coordinates $q^{\mu}$:

\begin{equation}
\label{eq:metric}
G_{\nu\rho}(q^{\mu}):=\sum_{\sigma=1}^{N} 
 \frac{\partial x^{\sigma}(q^{\mu})}{\partial q^{\nu}}
 m_{\sigma}\frac{\partial x^{\sigma}(q^{\mu})}{\partial q^{\rho}} \ ,
\end{equation}

which, in matrix form, reads

\begin{equation}
\label{eq:metric_matrix}
G=J^{T}MJ \ .
\end{equation}

The only difference with eq.~(\ref{eq:reduced_metric_matrix}) is that,
instead of the rectangular matrix $J_{c}$ (see eq.~(\ref{eq:Jc})),
in the above expression the full \emph{Jacobian matrix} of the change
of coordinates from Cartesian to curvilinear coordinates appears:

\begin{equation}
\label{eq:J}
J^{\sigma}_{\,\,\rho}(q^{\mu}):=
 \frac{\partial x^{\sigma}(q^{\mu})}{\partial q^{\rho}} \ .
\end{equation}

Obviously, one can deduce the factorization of $\det G$ as
a particular case of the results of sec.~\ref{sec:constrained}
with $L=0$, so that the indices $i,j$ now run over all
internal coordinates $q^{a}$. Explicitly,

\begin{equation}
\label{eq:detG_1}
\det G(q^{A},q^{a}) =
\sin^{2}\theta \, \det G_{2}(q^{a}) \ ,
\end{equation}

with

\begin{equation}
\label{eq:G2}
G_{2}:=\!
\left( \begin{array}{cc|c@{\hspace{2pt}}c@{\hspace{2pt}}c}
m_{tot}\displaystyle I^{(3)} & \displaystyle m_{tot}\,v(\vec{R}) &
\cdots & \displaystyle m_{tot} \frac{\partial \vec{R}}{\partial q^{b}} &
\cdots \\[10pt]
\displaystyle m_{tot}\,v^{T}(\vec{R}) & \mathcal{J} &
\cdots & \displaystyle \mathop{\Sigma}_{\alpha}m_{\alpha}
\frac{\partial \vec{x}^{\,\prime}_{\alpha}}{\partial q^{b}} \times
\vec{x}^{\,\prime}_{\alpha} & \cdots \\[10pt]
\hline
\vdots & \vdots & & \vdots & \\
\displaystyle m_{tot}\frac{\partial \vec{R}}{\partial q^{a}} &
\displaystyle \mathop{\Sigma}_{\alpha}m_{\alpha} \left(
\frac{\partial \vec{x}^{\,\prime}_{\alpha}}{\partial q^{a}} \times
\vec{x}^{\,\prime}_{\alpha} \right)^{T}&
\cdots &
\displaystyle \mathop{\Sigma}_{\alpha}m_{\alpha}
\frac{\partial \vec{x}^{\,\prime\,T}_{\alpha}}{\partial q^{a}}
\frac{\partial \vec{x}^{\,\prime}_{\alpha}}{\partial q^{b}} & \cdots \\[-2pt]
\vdots & \vdots & & \vdots & 
\end{array} \right) \ .
\end{equation}

However, in this section we would like to benefit
from the special structure of eq.~(\ref{eq:metric_matrix}), where,
differently from the constrained case, only $N \times N$ matrices
occur, and find an expression simpler than eq.~(\ref{eq:detG_1}).

If we take determinants on both sides of eq.~(\ref{eq:metric_matrix}),
we obtain

\begin{equation}
\label{eq:detGJ}
\det G = \left( \prod_{\alpha=1}^{n} m_{\alpha}^{3} \right) {\det}^{2} J \ .
\end{equation}

Similarly to eq.~(\ref{eq:Jc}), $J$ may be written
as follows:

\begin{equation}
\label{eq:J_explicit}
J=\left(
\begin{array}{cccc|c@{\hspace{2pt}}c@{\hspace{2pt}}c}
I^{(3)} & & 0 & & & 0 & \\[5pt]
\hline
& & & & & & \\[-10pt]
I^{(3)} & \displaystyle \frac{\partial E}{\partial \phi}\vec{x}^{\,\prime}_{2}
& \displaystyle \frac{\partial E}{\partial \theta}\vec{x}^{\,\prime}_{2}
& \displaystyle \frac{\partial E}{\partial \psi}\vec{x}^{\,\prime}_{2}
& \cdots &
\displaystyle E\frac{\partial \vec{x}^{\,\prime}_{2}}{\partial q^{b}} & 
\cdots \\[10pt]
I^{(3)} & \displaystyle \frac{\partial E}{\partial \phi}\vec{x}^{\,\prime}_{3}
& \displaystyle \frac{\partial E}{\partial \theta}\vec{x}^{\,\prime}_{3}
& \displaystyle \frac{\partial E}{\partial \psi}\vec{x}^{\,\prime}_{3}
& \cdots &
\displaystyle E\frac{\partial \vec{x}^{\,\prime}_{3}}{\partial q^{b}} & 
\cdots \\[-2pt]
\vdots & \vdots & \vdots & \vdots & & \vdots & \\
I^{(3)} & \displaystyle \frac{\partial E}{\partial \phi}
 \vec{x}^{\,\prime}_{\alpha}
& \displaystyle \frac{\partial E}{\partial \theta}\vec{x}^{\,\prime}_{\alpha}
& \displaystyle \frac{\partial E}{\partial \psi}\vec{x}^{\,\prime}_{\alpha}
& \cdots &
\displaystyle E\frac{\partial \vec{x}^{\,\prime}_{\alpha}}{\partial q^{b}} & 
\cdots \\[-2pt]
\vdots & \vdots & \vdots & \vdots & & \vdots &
\end{array}
\right)
\ .
\end{equation}

Now, the following identity is useful:

\begin{equation}
\label{eq:EJ}
J=
\left( \begin{array}{cccc}
I^{(3)} & & & 0 \\
 & E & & \\
 & & \ddots & \\
0 & & & E
\end{array} \right)J_{1}
\ ,
\end{equation}

where we have defined

\begin{equation}
\label{eq:J1}
J_{1}:=\left(
\begin{array}{cccc|c@{\hspace{2pt}}c@{\hspace{2pt}}c}
I^{(3)} & & 0 & & & 0 & \\[5pt]
\hline
& & & & & & \\[-10pt]
I^{(3)}& \displaystyle E^{T}
         \frac{\partial E}{\partial\phi}\vec{x}^{\,\prime}_{2}
& \displaystyle E^{T}\frac{\partial E}{\partial \theta}\vec{x}^{\,\prime}_{2}
& \displaystyle E^{T}\frac{\partial E}{\partial \psi}\vec{x}^{\,\prime}_{2}
& \cdots &
\displaystyle \frac{\partial \vec{x}^{\,\prime}_{2}}{\partial q^{b}} & 
\cdots \\[10pt]
I^{(3)} & \displaystyle E^{T}
          \frac{\partial E}{\partial \phi}\vec{x}^{\,\prime}_{3}
& \displaystyle E^{T}\frac{\partial E}{\partial \theta}\vec{x}^{\,\prime}_{3}
& \displaystyle E^{T}\frac{\partial E}{\partial \psi}\vec{x}^{\,\prime}_{3}
& \cdots &
\displaystyle \frac{\partial \vec{x}^{\,\prime}_{3}}{\partial q^{b}} & 
\cdots \\[-2pt]
\vdots & \vdots & \vdots & \vdots & & \vdots & \\
I^{(3)} & \displaystyle E^{T} \frac{\partial E}{\partial \phi}
 \vec{x}^{\,\prime}_{\alpha}
& \displaystyle E^{T}
                \frac{\partial E}{\partial \theta}\vec{x}^{\,\prime}_{\alpha}
& \displaystyle E^{T}
                \frac{\partial E}{\partial \psi}\vec{x}^{\,\prime}_{\alpha}
& \cdots &
\displaystyle \frac{\partial \vec{x}^{\,\prime}_{\alpha}}{\partial q^{b}} & 
\cdots \\[-2pt]
\vdots & \vdots & \vdots & \vdots & & \vdots &
\end{array}
\right)
\ ,
\end{equation}

and we have that only the determinant of $J_{1}$ needs to be
computed, since $\det J = {\det}^{n-1}E\,\det J_{1} = \det J_{1}$.

Next, we note that, according to the definition of the primed
reference frame in sec.~\ref{sec:definitions}, some of the components
of the vectors $\vec{x}^{\,\prime}_{2}$ and $\vec{x}^{\,\prime}_{3}$
are zero, namely, we have that

\begin{equation}
\label{eq:x2x3}
\vec{x}^{\,\prime}_{2} = \left( \begin{array}{c} 0 \\ 0 \\
  x^{\,\prime\,3}_{2} \end{array}\right) \quad \mathrm{and} \quad
\vec{x}^{\,\prime}_{3} = \left( \begin{array}{c} x^{\,\prime\,1}_{3} \\ 0 \\
  x^{\,\prime\,3}_{3} \end{array}\right) \ .
\end{equation}

Hence, the derivatives with respect to $q^{b}$ of the zero components
are also zero, rendering three zero rows in the bottom right block of
eq.~(\ref{eq:J1}). Performing two row permutations so that the zero rows
are the top-most ones, we obtain a matrix $J_{2}$ whose determinant is
the same as the one of $J_{1}$:

\begin{equation}
\label{eq:J2}
J_{2}=
\left( \begin{array}{ccc}
I^{(3)} & & 0 \\
 & J_{2}^{\mathcal{E}} & \\
\mathcal{X} & & J_{2}^{\mathcal{I}}
\end{array} \right)
\ ,
\end{equation}

where the blocks in the diagonal have been defined as

\begin{equation}
\label{eq:J2_E}
J_{2}^{\mathcal{E}}=
\left( \begin{array}{ccc}
\displaystyle \left(E^{T}\frac{\partial E}{\partial \phi}
              \vec{x}^{\,\prime}_{2} \right)^{1} &
\displaystyle \left(E^{T}\frac{\partial E}{\partial \theta}
              \vec{x}^{\,\prime}_{2} \right)^{1} &
\displaystyle \left(E^{T}\frac{\partial E}{\partial \psi}
              \vec{x}^{\,\prime}_{2} \right)^{1} \\[10pt]
\displaystyle \left(E^{T}\frac{\partial E}{\partial \phi}
              \vec{x}^{\,\prime}_{2} \right)^{2} &
\displaystyle \left(E^{T}\frac{\partial E}{\partial \theta}
              \vec{x}^{\,\prime}_{2} \right)^{2} &
\displaystyle \left(E^{T}\frac{\partial E}{\partial \psi}
              \vec{x}^{\,\prime}_{2} \right)^{2} \\[10pt]
\displaystyle \left(E^{T}\frac{\partial E}{\partial \phi}
              \vec{x}^{\,\prime}_{3} \right)^{2} &
\displaystyle \left(E^{T}\frac{\partial E}{\partial \theta}
              \vec{x}^{\,\prime}_{3} \right)^{2} &
\displaystyle \left(E^{T}\frac{\partial E}{\partial \psi}
              \vec{x}^{\,\prime}_{3} \right)^{2} \\
\end{array} \right)
\ ,
\end{equation}

the superindices standing for vector components, and

\begin{equation}
\label{eq:J2_I}
J_{2}^{\mathcal{I}}=
\left( \begin{array}{ccc}
\cdots &
\displaystyle \frac{\partial x^{\,\prime\,3}_{2}}{\partial q^{b}} &
\cdots \\[10pt]
\cdots &
\displaystyle \frac{\partial x^{\,\prime\,1}_{3}}{\partial q^{b}} &
\cdots \\[10pt]
\cdots &
\displaystyle \frac{\partial x^{\,\prime\,3}_{3}}{\partial q^{b}} &
\cdots \\[10pt]
\cdots &
\displaystyle \frac{\partial \vec{x}^{\,\prime}_{4}}{\partial q^{b}} &
\cdots \\
 & \vdots & \\
\cdots &
\displaystyle \frac{\partial \vec{x}^{\,\prime}_{\alpha}}{\partial q^{b}} &
\cdots \\
 & \vdots &
\end{array} \right) \ .
\end{equation}

The concrete form of the submatrix $\mathcal{X}$ in eq.~(\ref{eq:J2}) is
irrelevant for our purposes, since

\begin{equation}
\label{eq:detJdetJ2}
\det J = \det J_{2} = \det J_{2}^{\mathcal{E}} \det J_{2}^{\mathcal{I}} \ .
\end{equation}

An explicit computation of $J_{2}^{\mathcal{E}}$ yields

\begin{equation}
\label{eq:J2_E_explicit}
J_{2}^{\mathcal{E}}=
\left( \begin{array}{ccc}
x^{\,\prime\,3}_{2}\sin\theta\sin\psi & -x^{\,\prime\,3}_{2}\cos\psi & 0 \\
x^{\,\prime\,3}_{2}\sin\theta\cos\psi & x^{\,\prime\,3}_{2}\sin\psi & 0 \\
-x^{\,\prime\,1}_{3}\cos\theta + x^{\,\prime\,3}_{3}\sin\theta\cos\psi &
x^{\,\prime\,3}_{3}\sin\psi & x^{\,\prime\,1}_{3}
\end{array} \right) \ ,
\end{equation}

with determinant $\det J_{2}^{\mathcal{E}} =
\sin\theta\,x^{\,\prime\,1}_{3} \bigl(x^{\,\prime\,3}_{2}\bigr)^{2}$.

Using eqs.~(\ref{eq:detJdetJ2}) and (\ref{eq:detGJ}), we finally obtain

\begin{equation}
\label{eq:detG_final}
\det G(q^{A},q^{a}) =
         {\sin}^{2}\theta \bigl(x^{\,\prime\,1}_{3}(q^{a})\bigr)^{2}
         \bigl(x^{\,\prime\,3}_{2}(q^{a})\bigr)^{4}
         \det J_{2}^{\mathcal{I}}(q^{a})
\left( \prod_{\alpha=1}^{n} m_{\alpha}^{3} \right) \ ,
\end{equation}

where the factorization has been achieved, since the
only factor that depends on the external coordinates is
${\sin}^{2}\theta$.

\section{Conclusions}
\label{sec:conclusions}

In this work, we have calculated explicit expressions in which the
determinant of the mass-metric tensor $G$ (eqs.~(\ref{eq:detG_final})
and (\ref{eq:detG_1})) and the determinant of the reduced mass-metric
tensor $g$ (eq.~(\ref{eq:detg})), occurring in Classical Statistical
Mechanics in the coordinate space, are written as a product of two
functions; one depending only on the external coordinates that
describe the overall translation and rotation of the system, and the
other only on the internal coordinates. This has been done for any
molecule, general internal coordinates and arbitrary constraints,
extending the work in refs.~\citen{PE:Go1976MM} and
\citen{PE:Pat2004JCP}.

This factorization allows to integrate out the external coordinates
and perform Monte Carlo simulations in the internal conformational
space, gaining insight of the problem, simplicity in the description
of the system and, for small molecules, some computational
effort. Also, our results indicate that, in general, the Fixman's
compensating potential
\cite{PE:Fix1974PNAS,PE:Ech2006JCCb,PE:Mor2004ACP,PE:Den2000MP},
customarily used to reproduce the stiff equilibrium distribution using
rigid molecular dynamics simulations, does not depend on the external
variables.

In appendix A, we give a general mathematical argument showing that
the factorization is a consequence of the symmetries of the metric
tensors involved and, in appendix B, the determinant of the
mass-metric tensor $G$ is computed explicitly in the SASMIC
\cite{PE:Ech2006JCCa} set of curvilinear coordinates for general
branched molecules (see eq.~(\ref{eq:detG_sasmic})) showing that the
classical formula for serial polymers \cite{PE:Go1976MM} is actually
valid for any macromolecule.

All the expressions derived in the present work are directly
applicable to real cases, as has been checked in
ref.~\citen{PE:Ech2006JCCb}.

\section*{Acknowledgments}

\hspace{0.5cm} We would like to thank J. L. Alonso and F. Falceto for
illuminating discussions.

This work has been supported by the Arag\'on
Government (``Biocomputaci\'on y F\'{\i}sica de Sistemas Complejos''
group) and by the research grants MEC (Spain) \mbox{FIS2004-05073} and
\mbox{FPA2003-02948}. P. Echenique and I. Calvo are supported by MEC
(Spain) FPU grants.

\appendix

\setcounter{secnumdepth}{-1}
\section{Appendix A:\\
{\large General mathematical argument}}
\label{sec:appendixA}
\setcounter{equation}{0}
\setcounter{secnumdepth}{2}
\setcounter{section}{1}

Let ${\cal M}$ be a finite dimensional differentiable manifold
equipped with a riemannian metric tensor. Take local coordinates
$q^{\mu}$ on ${\cal M}$ and denote by $G_{\mu\nu}(q)$ the components of
the metric tensor in these coordinates.

The transformation

\begin{equation}
\label{eq:inf_transf}
q^{\,\prime\mu} = q^{\mu} + \epsilon\,\xi^{\mu}(q) + O(\epsilon^2)
\end{equation}

is said an \emph{isometry} and $\xi^{\mu}(x)$ is said a
\emph{Killing vector field} if

\begin{equation}
\label{eq:isometry}
G_{\mu\nu}\big(q^{\,\prime}(q)\big)=J^{\rho}_{\ \mu}\big(q^{\,\prime}(q)\big)\,
                            G_{\rho\sigma}(q)\,
                            J^{\sigma}_{\ \nu}\big(q^{\,\prime}(q)\big) \ ,
\end{equation}

where

\begin{equation}
\label{eq:J_transf}
J^{\mu}_{\ \nu}\big(q^{\,\prime}(q)\big):=\left( \frac{\partial q^{\mu}}
          {\partial q^{\,\prime\nu}}\right)\big(q^{\,\prime}(q)\big) \ .
\end{equation}

Expanding eq.~(\ref{eq:isometry} up to first order in $\epsilon$ and
noticing that ${\rm det}(J^{\mu}_{\ \nu}) = 1 - \epsilon \,
\partial_{\mu}\xi^{\mu}(q)$, we obtain the following differential
equation for $\mathcal{G}:={\rm det}(G_{\mu\nu})$:

\begin{equation}
\label{eq:diffeq}
\xi^\mu(q)\partial_{\mu} \mathcal{G}(q) =
  -2 \big(\partial_{\mu}\xi^{\mu}(q)\big)\,
 \mathcal{G}(q) \ .
\end{equation}

Let us apply this machinery to the case considered in this work. For
concreteness, we shall derive the factorization of the external
coordinates in the unconstrained case and shall argue that this still
holds in the constrained one.

Simultaneous translations and rotations of all the
particles\footnote{Notice that the isometry group of the mass-metric
tensor is much bigger, since translations and rotations acting
independently on each particle are also isometry transformations.} are
isometries of the mass-matrix tensor in eq.~(\ref{eq:mass_matrix}). The
important point for us is that, in the coordinates $q^{\mu}$ introduced
in sec.~\ref{sec:definitions}, these transformations change the
external coordinates $(X,Y,Z,\phi,\theta,\psi)$ and leave the internal
coordinates $q^{a}$ untouched (see eq.~(\ref{eq:transf1_unconstrained})).

A global translation is given in Cartesian coordinates by
$x_{\alpha}^{p} \mapsto x_{\alpha}^{p}+\epsilon$. In the coordinates
$q^{\mu}$, it takes $(X,Y,Z)\mapsto (X,Y,Z)+\epsilon\,(1,1,1)$ and does
not affect the remaining coordinates. With the above notation,
$\xi^{\mu} = 1,\ \mu=1,2,3$ and $\xi^{\mu}=0,\ \forall\mu > 3$. Hence,
eq.~(\ref{eq:diffeq}) implies that

\begin{equation}
\label{eq:Xindep}
\partial_{X}\mathcal{G} = \partial_{Y}\mathcal{G} =
 \partial_{Z}\mathcal{G} = 0 \ ,
\end{equation}

i.e., the determinant of the mass-metric tensor does not depend on the
coordinates $X,Y,Z$.

A global rotation in the coordinates $q^{\mu}$ rotates $(X,Y,Z)$ and
changes the Euler angles (in a complicated way which will not be
important for our purposes) but does not affect the internal
coordinates. Hence, $\xi^{\mu} = 0,\forall \mu > 6$. In addition, the
matrix $J^{\mu}_{\ \nu}$ does not depend on $X,Y,Z$ because the
rotation acts linearly on them. Let us abbreviate $\alpha \equiv
\alpha^{p} \equiv (\phi,\theta,\psi),\ p=1,\ldots,3$. Recalling that
$\mathcal{G}$ does not depend on $X,Y,Z$, the differential
equation~(\ref{eq:diffeq}) reads

\begin{equation}
\label{eq:factoreq}
\xi^{p}(\alpha) \partial_{p}\mathcal{G}(\alpha,q^{a}) = 
 -2(\partial_{p}\xi^{p}(\alpha))\,\mathcal{G}(\alpha,q^{a}) \ .
\end{equation}

The group of rotations in ${\mathbb R}^3$ has three linearly
independent Killing vector fields which are complete in the sense that
one can join two arbitrary points $(\phi,\theta,\psi)$ and
$(\phi^{\,\prime},\theta^{\,\prime},\psi^{\,\prime})$ by moving along
integral curves of the Killing vector fields. This guarantees that the
solution of eq.~(\ref{eq:factoreq}) is of the form

\begin{equation}
\label{eq:final_arg}
\mathcal{G}(\alpha,q^{a})=\mathcal{G}_{1}(\alpha)\,\mathcal{G}_{2}(q^{a})
\end{equation}

and we have the desired result.

To derive the factorization of the external coordinates in the
constrained case, simply notice that the constraints in this work do
not involve the external coordinates. Therefore, global translations and
rotations are still isometries of the reduced mass-metric tensor and the
result follows.

\setcounter{secnumdepth}{-1}
\section{Appendix B:\\
{\large Determinant of G in particular coordinates}}
\label{sec:appendixB}
\setcounter{equation}{0}
\setcounter{secnumdepth}{2}

The SASMIC scheme, introduced in ref.~\citen{PE:Ech2006JCCa}, is
a set of rules to define particular Z-matrix coordinates
\cite{PE:Gaussian70,PE:Lev1999BOOK} of general branched molecules,
with convenient properties of modularity and approximate separability
of soft and hard modes.

According to the rules, to each atom $\alpha$, one uniquely
assigns three atoms $\beta (\alpha )$, $\gamma (\alpha )$ and
$\delta (\alpha )$ in such a way that the three Z-matrix
internal coordinates that position atom $\alpha$ are

\begin{equation}
\label{eq:sasmic}
\begin{array}{l}
r_{\alpha}:=\big(\alpha,\beta (\alpha )\big) \\
\theta_{\alpha}:=\big(\alpha,\beta (\alpha ),\gamma (\alpha )\big) \\
\phi_{\alpha}:=\big(\alpha,\beta (\alpha ),\gamma (\alpha ),
                    \delta (\alpha )\big) \ ,
\end{array}
\end{equation}

being $r_{\alpha}$ a bond length, $\theta_{\alpha}$ a bond angle and
$\phi_{\alpha}$ a dihedral angle.

\begin{figure}[!ht]
\begin{center}
\epsfig{file=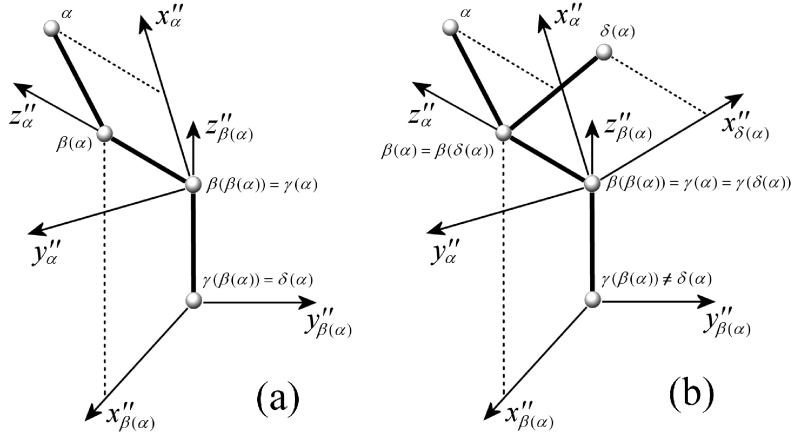,width=11cm}
\end{center}
\caption{\label{fig:axes2}{\small Local reference frames associated to
atoms $\alpha$ and $\beta(\alpha )$ (see text) in the cases that {\bf
(a)} $\phi_{\alpha}$ is a principal dihedral or {\bf (b)}
$\phi_{\alpha}$ is a phase dihedral.}}
\end{figure}

The procedure that will be followed in order to express the position
$\vec{x}^{\,\prime}_{\alpha}$ of atom $\alpha$ in the primed reference
frame in fig.~\ref{fig:axes} as a function of the SASMIC internal
coordinates starts by expressing the vector that goes from $\beta
(\alpha )$ to $\alpha$ in a set of axes
$(x^{\,\prime\prime}_{\alpha},y^{\,\prime\prime}_{\alpha},z^{\,\prime\prime}_{\alpha})$
associated to $\alpha$. This local reference frame is defined such
that the $z^{\,\prime\prime}_{\alpha}$-axis lies along the directional
bond $\big(\gamma (\alpha ),\beta (\alpha )\big)$ and the
$x^{\,\prime\prime}_{\alpha}$-axis lies along the projection of
$\big(\beta (\alpha ),\alpha\big)$ onto the plane orthogonal to
$\big(\gamma (\alpha ),\beta (\alpha )\big)$ (see
fig.~\ref{fig:axes2}).

In these axes, the components of the vector $\big(\beta
(\alpha ),\alpha\big)$ are

\begin{equation}
\label{eq:xpp}
\vec{x}^{\,\prime\prime\,T}_{\alpha}:=(r_{\alpha}\sin\theta_{\alpha} , 0 ,
                                       -r_{\alpha}\cos\theta_{\alpha}) \ .
\end{equation}

Now, if the atom $\delta ( \alpha )$ that is used to define
$\phi_{\alpha}$ is bonded to atom $\gamma ( \alpha )$
(fig.~\ref{fig:axes2}a), $\phi_{\alpha}$ is called a
\emph{principal dihedral} \cite{PE:Ech2006JCCa} and we have that

\begin{equation}
\label{eq:rotate_xpp}
 \underbrace{
 \left( \begin{array}{ccc}
 -\cos\theta_{\beta(\alpha )} & 0 & \sin\theta_{\beta(\alpha )} \\
 0 & 1 & 0 \\
 -\sin\theta_{\beta(\alpha )} & 0 & -\cos\theta_{\beta(\alpha )}
 \end{array} \right)
 }_{\displaystyle \Theta(\theta_{\beta(\alpha )})}
 \underbrace{
 \left( \begin{array}{ccc}
 \cos\phi_{\alpha} & -\sin\phi_{\alpha} & 0 \\
 \sin\phi_{\alpha} & \cos\phi_{\alpha} & 0 \\
 0 & 0 & 1
 \end{array} \right)
 }_{\displaystyle \Phi(\phi_{\alpha})}
 \vec{x}^{\,\prime\prime}_{\alpha}
\end{equation}

are the components of the vector $\big(\beta (\alpha ),\alpha\big)$ in
the local reference frame $(x^{\,\prime\prime}_{\beta(\alpha
)},y^{\,\prime\prime}_{\beta(\alpha
)},z^{\,\prime\prime}_{\beta(\alpha )})$ associated to atom
$\beta(\alpha )$.

On the other hand, if we are at a branching point and the atom $\delta
( \alpha )$ that is used to define $\phi_{\alpha}$ is bonded to atom
$\beta ( \alpha )$ (fig.~\ref{fig:axes2}b), $\phi_{\alpha}$ is called
a \emph{phase dihedral} \cite{PE:Ech2006JCCa} and we have to change
first to the local reference frame associated to $\delta ( \alpha )$.
In this case, the components of the vector $\big(\beta (\alpha
),\alpha\big)$ in the local reference frame
$(x^{\,\prime\prime}_{\beta(\alpha )},y^{\,\prime\prime}_{\beta(\alpha
)},z^{\,\prime\prime}_{\beta(\alpha )})$ are

\begin{equation}
\label{eq:rotate_xpp_phase}
 \Theta(\theta_{\beta(\alpha )})\Phi(\phi_{\delta(\alpha)})\Phi(\phi_{\alpha})
 \,\vec{x}^{\,\prime\prime}_{\alpha} \ .
\end{equation}

If we iterate the procedure, by changing the axes to the ones
associated to the atom $\beta(\beta(\alpha))$, i.e., the $\beta$ atom
that correspond to $\beta(\alpha)$ according to the SASMIC scheme, an
so on, we will eventually arrive to the set of axis
$(x^{\,\prime\prime}_{3},y^{\,\prime\prime}_{3},z^{\,\prime\prime}_{3})$
(since, in the SASMIC scheme \cite{PE:Ech2006JCCa}, we have that
$\beta(\alpha)<\alpha$). Note however that, according to the
definition of the local reference frame given in the preceding
paragraphs, the one associated to atom 3 is \emph{exactly} the primed
reference frame in fig.~\ref{fig:axes}.

Hence, let us define, for each atom $\alpha$, a matrix $R_{\alpha}$ as the
product of the matrices obtained using eqs.~(\ref{eq:rotate_xpp}) and
(\ref{eq:rotate_xpp_phase}) and successively applying the function
$\beta(\alpha)$. Then, $R_{\alpha}$ takes the vector $\big(\beta
(\alpha),\alpha\big)$ in eq.~(\ref{eq:xpp}) to the primed reference
frame.

Let the superindex on $\beta$ denote composition of functions, let us
define $\beta^{0}(\alpha):=\alpha$ and $\mathcal{N}_{\alpha}$ as the
integer such that $\beta^{\mathcal{N}_{\alpha}+1}(\alpha)=3$.  Adding
all the vectors corresponding to $\big(\beta^{p+1} (\alpha),\beta^{p}
(\alpha)\big)$ in the primed reference frame, with
$p=0,\ldots,\mathcal{N}_{\alpha}$, to $\vec{x}^{\,\prime}_{3}$ yields
the position of atom $\alpha$ in the primed reference frame as a
function of the internal coordinates:

\begin{equation}
\label{eq:xp}
 \vec{x}^{\,\prime}_{\alpha} = \vec{x}^{\,\prime}_{3} +
 \sum_{p=\mathcal{N}_{\alpha}}^{0}
 R_{\beta^{p}(\alpha)}\vec{x}^{\,\prime\prime}_{\beta^{p}(\alpha)} \ .
\end{equation}

Now, ordering the internal coordinates as $(r_{2}, r_{3},
\theta_{3}, r_{4}, \theta_{4}, \phi_{4}, \ldots,r_{n}, \theta_{n},
\phi_{n})$ and using the already mentioned fact that
$\beta(\alpha)<\alpha$, and also that $\delta(\alpha)<\alpha$, we have
that the matrix $J_{2}^{\mathcal{I}}$ in eq.~(\ref{eq:J2_I}) is

\begin{equation}
\label{eq:J2_I_tri}
J_{2}^{\mathcal{I}}=
\left( \begin{array}{cccc}
A_{0} & & & 0 \\
 & A_{4} & & \\
 & & \ddots & \\
\mathcal{X} & & & A_{n}
\end{array} \right) \quad \mathrm{and} \quad
 \det J_{2}^{\mathcal{I}}=\det A_{0}\prod_{\alpha=4}^{n}\det A_{\alpha} \ .
\end{equation}

Using that

\begin{equation}
\label{eq:xp2_xp3}
\vec{x}^{\,\prime}_{2}=
\left( \begin{array}{c} 0 \\ 0 \\ r_{2} \end{array} \right) \quad
\mathrm{and} \quad
\vec{x}^{\,\prime}_{3}=
\left( \begin{array}{c}
r_{3}\sin\theta_{3} \\
0 \\
r_{2} - r_{3}\cos\theta_{3}
\end{array} \right)
\ ,
\end{equation}

we have

\begin{equation}
\label{eq:A0}
A_{0}:=
\left( \begin{array}{ccc}
\displaystyle \frac{\partial x^{\,\prime\,3}_{2}}{\partial r_{2}} &
\displaystyle \frac{\partial x^{\,\prime\,3}_{2}}{\partial r_{3}} &
\displaystyle \frac{\partial x^{\,\prime\,3}_{2}}{\partial \theta_{3}} \\[10pt]
\displaystyle \frac{\partial x^{\,\prime\,1}_{3}}{\partial r_{2}} &
\displaystyle \frac{\partial x^{\,\prime\,1}_{3}}{\partial r_{3}} &
\displaystyle \frac{\partial x^{\,\prime\,1}_{3}}{\partial \theta_{3}} \\[10pt]
\displaystyle \frac{\partial x^{\,\prime\,3}_{3}}{\partial r_{2}} &
\displaystyle \frac{\partial x^{\,\prime\,3}_{3}}{\partial r_{3}} &
\displaystyle \frac{\partial x^{\,\prime\,3}_{3}}{\partial \theta_{3}}
\end{array} \right) =
\left( \begin{array}{ccc}
1 & 0 & 0 \\[10pt]
0 & \sin\theta_{3} & r_{3}\cos\theta_{3} \\[10pt]
1 & -\cos\theta_{3} & r_{3}\sin\theta_{3}
\end{array} \right)
\ .
\end{equation}

Now, we note that the matrix $\Phi(\phi_{\alpha})$ occurs always
at the right-most place in $R_{\alpha}$ and that the derivatives
in the blocks $A_{\alpha}$, with $\alpha > 4$, kill all the terms
in eq.~(\ref{eq:xp}) except for the one corresponding to $p=0$. Hence,
if we define $R_{\alpha}=:R^{\,\prime}_{\alpha}\,\Phi(\phi_{\alpha})$,
the block $A_{\alpha}$ may be expressed as follows:

\begin{equation}
\label{eq:Aalpha}
\begin{array}{c}
\displaystyle
A_{\alpha} := R^{\,\prime}_{\alpha}
\left(  \frac{\partial
  \Phi(\phi_{\alpha})\vec{x}^{\,\prime\prime}_{\alpha}}
  {\partial r_{\alpha}} \ \
  \frac{\partial \Phi(\phi_{\alpha})\vec{x}^{\,\prime\prime}_{\alpha}}
  {\partial \theta_{\alpha}} \ \
  \frac{\partial \Phi(\phi_{\alpha})\vec{x}^{\,\prime\prime}_{\alpha}}
  {\partial \phi_{\alpha}} \right) = \\[14pt]
\displaystyle \quad
\left( \begin{array}{ccc}
\sin\theta_{\alpha}\cos\phi_{\alpha} &
r_{\alpha}\cos\theta_{\alpha}\cos\phi_{\alpha} &
-r_{\alpha}\sin\theta_{\alpha}\sin\phi_{\alpha} \\
\sin\theta_{\alpha}\sin\phi_{\alpha} &
r_{\alpha}\cos\theta_{\alpha}\sin\phi_{\alpha} &
r_{\alpha}\sin\theta_{\alpha}\cos\phi_{\alpha} \\
-\cos\theta_{\alpha} & r_{\alpha}\sin\theta_{\alpha} & 0
\end{array} \right) \ .
\end{array}
\end{equation}

Finally, using eq.~(\ref{eq:J2_I_tri}), noting that $\det A_{0} =
r_{3}$ and $\det A_{\alpha} = -r_{\alpha}^{2}\sin\theta_{\alpha}$, and
calculating the remaining terms of eq.~(\ref{eq:detG_final}) with
eq.~(\ref{eq:xp2_xp3}), we obtain the desired result:

\begin{equation}
\label{eq:detG_sasmic}
{\det}^{\frac{1}{2}}G(q^{A},q^{a}) =
         \left( \prod_{\alpha=1}^{n} m_{\alpha}^{3/2} \right)
         |\sin\theta| \left( \prod_{\alpha=2}^{n} r_{\alpha}^{2} \right)
          \left( \prod_{\alpha=3}^{n} |\sin\theta_{\alpha}| \right) \ .
\end{equation}

It is worth remarking at this point that the previous expression does
not explicitly depend on the dihedral angles $\phi_{\alpha}$ and that
it is the same result as the one found in
ref.~\citen{PE:Go1976MM} for serial polymers.

\bibliography{externals}

\end{document}